\documentclass[letterpaper,twocolumn,10pt]{article}

\usepackage[latin1]{inputenc}
\usepackage[arabic,english]{babel}
\usepackage{scs}
\usepackage{graphicx}
\usepackage{citesort}
\usepackage{amssymb,amsmath}
\usepackage{rotating}
\usepackage{natbib}
\usepackage{hyperref,epstopdf,arabtex}
\begin{document}
\setarab
\vocalize
%
%
\title{Solving the Klein-Gordon equation using Fourier spectral methods: A benchmark test for computer performance}
\author{S. Aseeri\footnotemark[1], O. Batra\u{s}ev$^2$, M. Icardi$^3$, B. Leu$^4$, A. Liu$^4$, N. Li$^5$, B.K. Muite$^{2,6}$, E. M\"{u}ller$^7$, B. Palen$^8$, M. Quell$^{9}$, H. Servat$^{10}$, P. Sheth$^{11}$, R. Speck$^{12}$, M. Van Moer$^{13}$, J. Vienne$^{14}$} 

\maketitle


\begin{abstract}
\noindent
The cubic Klein-Gordon equation is a simple but non-trivial partial differential equation whose numerical solution has the main building blocks required for the solution of many other partial differential equations. In this study, the library 2DECOMP\&FFT is used in a Fourier spectral scheme to solve the Klein-Gordon equation and strong scaling of the code is examined on thirteen different machines for a problem size of $512^3$. The results are useful in assessing likely performance of other parallel fast Fourier transform based programs for solving partial differential equations.  The problem is chosen to be large enough to solve on a workstation, yet also of interest to solve quickly on a supercomputer, in particular for parametric studies. Unlike other high performance computing benchmarks, for this problem size, the time to solution will not be improved by simply building a bigger supercomputer.
\end{abstract}

%
%
\section{Introduction}
The focusing nonlinear Klein-Gordon equation describes the evolution of a possibly complex scalar field $u$ according to\footnotetext[1]{KAUST Supercomputing Laboratory, 4700 King Abdullah University of Science and Technology,Thuwal 23955-6900, Kingdom of Saudi Arabia, 
$^2$Arvutiteaduse instituut, Tartu \"{U}likool, Liivi 2, Tartu 50409, Estonia, 
$^3$Stochastic Numerics Group, 4700 King Abdullah University of Science and Technology, Thuwal 23955-6900, Kingdom of Saudi Arabia, 
$^4$Electrical and Computer Engineering, University of Michigan, Ann Arbor, MI  48109-2122, USA, 
$^5$Numerical Algorithms Group, Wilkinson House, Jordan Hill Road, Oxford, OX2 8DR, United Kingdom, 
$^6$\RL{Als`wdyT} \RL{Al`rbyT} \RL{AlmamlkT} \RL{6900-23955,} \RL{_twl}  \RL{lel`lwm wAltqnyT,}  \RL{`bdAlalh} \RL{Almlk}  \RL{jAm`T} \RL{4700}  \RL{Al`dadyT,} \RL{AlryA.dyAt} \RL{mjmw`T} (Numerical Mathematics Group, 4700 King Abdullah University of Science and Technology, Thuwal 23955-6900, Kingdom of Saudi Arabia), $^7$Department of Mathematical Sciences, University of Bath, Bath BA2 7AY, United Kingdom, 
$^8$CAEN Advanced Computing, University of Michigan, Ann Arbor, MI 48109-2094, USA, 
$^9$Institut f\"{u}r Analysis und Scientific Computing, Technische Universit\"{a}t Wien, Wiedner Haupstrasse 8-10, 1040 Wien, Austria, 
$^{10}$Computer Sciences Department, BSC, Department of Computer Architecture, UPC, c/Jordi Girona, 31 - 08034 - Barcelona, Catalunya, Spain, 
$^{11}$Biomedical Engineering, University of Michigan, Ann Arbor, MI  48109-2122, USA, 
$^{12}$Juelich Supercomputing Centre, Forschungszentrum Juelich GmbH, Germany, 
$^{13}$National Center for Supercomputing Applications, University of Illinois at Urbana-Champaign, Urbana, IL 61801, USA, 
$^{14}$Texas Advanced Computing Center, University of Texas at Austin, Austin, Texas 78758-4497, USA.} 
\begin{equation}\label{eq:KleinGordon}
u_{tt} - \Delta u +u = \lvert u\rvert^2u,
\end{equation}
where $t$ is time and $\Delta=\partial_{xx}+\partial_{yy}+\partial_{zz}$ the three-dimensional Laplacian. In this note we will focus on real valued functions $u$ only. This equation can exhibit a phenomena known as blow up in which the field $u$ becomes infinite at a certain point. Many partial differential equations which model physical phenomena can exhibit blow up, which typically indicates that the differential equation model is no longer a good model for the physical phenomenon under investigation. Understanding under which initial conditions a blow up in the Klein-Gordon equation will occur is still not clear and it is hoped that parametric simulations can help to elucidate this. Some parametric numerical studies of three dimensional radially symmetric real solutions to the Klein-Gordon equation have recently been done by \citet{DonSch11}. Their parametric study generated a large amount of data, which was unfeasible to store. Their study also made an assumption of spherical symmetry to be able to solve a one-dimensional equation for the radial component of a three-dimensional field. It is of interest to perform similar parametric simulations in three dimensions without symmetry assumptions for the Klein-Gordon and other partial differential equations. Fourier spectral methods are an effective tool for doing this on petascale supercomputers. The purpose of this paper is to conduct strong scaling experiments of a full model problem on different platforms, trying to understand how much faster a moderate size problem can be made to run. The Klein-Gordon equation is a simple model application for which different choices of numerical methods and computer architectures can be tried to determine which one will give an accurate enough solution at either the fastest time, lowest computational or lowest energy cost.  To limit the scope of this study, parallel fast Fourier transforms (FFT) are done with the library 2DECOMP\&FFT \citep{LiLai10}\citep{2DECOMP&FFT} which primarily uses MPI\_ALL\_TO\_ALLV or MPI\_ALL\_TO\_ALL for its communications. However, it would be interesting to perform a similar study with other parallel Fourier transform libraries (for example PFFT \citep{Pip13}\citep{PFFT}, OpenFFT \citep{DuyOza14}\citep{OpenFFT}, PKUFFT \citep{CheCuiMei10}\citep{PKUFFT} and P3DFFT \citep{Pek12}\citep{P3DFFT}). The current focus of the study is on a discretization of $512^3$ since (i) there are still a wide variety of numerical experiments that can be done for problems of this size, (ii) post processing can be done locally on workstations and (iii) for parametric studies, it is unclear whether access to high-performance computers or a distributed computing cloud based solution is most appropriate. It would also be interesting to try out other numerical methods, such as higher-order time stepping, a lattice-Boltzmann method \citep{LiJiZheLiu11} and an implicit finite difference/finite element spatial discretization approach with multigrid, fast multipole, tree code and/or preconditioned conjugate gradient solvers. For related work see \citet{GhoEtAl14}, from which it is clear that a careful choice of benchmarking solution is needed when comparing different discretization methods and elliptic system solvers. A later study will describe how such model could be used to rank computing systems as a possible addition to the Linpack benchmark \citep{Top500} and as an update of the NAS parallel benchmarks \citep{Bai11}. There are already several attempts to generate benchmarks to supplement the Linpack ranking of supercomputers, such as high performance conjugate gradient \citep{HPCG} and high performance multigrid \citep{HPGMG}. However, neither of these solve a real world problem or allow for algorithmic improvements in methods for solving differential equations and linear systems (see for example \citep{BalEtAl14} and for Poisson's equation \citep[p. 277]{Dem97}, \citep[Lecture21]{Dem14}). This paper aims to start a discussion on a simple method to rank supercomputers for solving realistic problems that also allows for algorithmic improvements, emphasizes the entire computer eco-system, including software that can be used and developed for that system, and the people that operate the system. \citet{RamVieWijKoeSha13} have already noted that better ways to evaluate accelerators are required. By stating a more general problem than solution of a linear system of equations, one can use the best architecture specific algorithm on each platform thereby making a computer ranking not only a measurement of CPU double precision floating point power, but also of problem solving effectiveness.

\subsection{The Klein-Gordon Equation}
\citet{NakSch11} give an introduction to some of the theory of the Klein-Gordon equation, focusing primarily on the three-dimensional radial case. Two-dimensional simulations of the Klein-Gordon equation can be found in \citet{BaoYan07} and \citet{Yan06}. The linear Klein-Gordon equation occurs as a modification of the linear Schr\"{o}dinger equation that is consistent with special relativity, see for example \citet{Lan96} or \citet{Gre94}. At the present time, there have been no numerical studies of blow up of solutions to this equation without the assumption of radial symmetry. This equation has generated a large mathematical literature and yet is still poorly understood. Most of this mathematical literature has focused on analyzing the equation on an infinite three-dimensional space with initial data that either decays exponentially as one tends to infinity or is nonzero on a finite set of the domain. Here, we will simulate this equation in a periodic setting. Since this equation is a wave equation, it has a finite speed of propagation of information, much as a sound wave in air takes time to move from one point to another. Consequently for short time simulations, a simulation of a solution that is only nonzero on a finite part of the domain is similar to a simulation on an infinite domain. However, over long times, the solution can spread out and interact with itself on a periodic domain, whereas on an infinite domain,  the interaction over long times is significantly reduced and the solution primarily spreads out. Understanding the interactions in a periodic setting is an interesting mathematical problem. Sufficiently smooth solutions of the Klein-Gordon equation conserve the energy given by
\begin{align*} 
E(u,u_t)=\int  \frac{1}{2}\lvert u_t\rvert^2 + \frac{1}{2}\lvert u\rvert^2+\frac{1}{2}\left\lvert \nabla u \right\rvert^2 - \frac{1}{4}\left\lvert u \right\rvert^4 \mathrm{d}\boldmath{ x}.
\end{align*}
When accurate time stepping schemes are used for sufficiently bounded and differentiable solutions, the energy can act as a test of the correctness of the code implementation, the libraries used and of the computer hardware, since if the energy is not conserved and the implementation is correct, there is likely to be a hardware or library error.

\subsubsection{Numerical Schemes}
The two time stepping schemes that were used by \citet{DonSch11} for radially symmetric solutions, can be readily adapted to fully three-dimensional simulations using Fourier pseudospectral discretization instead of a finite difference spatial discretization. One of these schemes is simple to implement, and a modification of this for implicit time stepping is described below since it is typical of numerical methods used for finding approximate solutions to partial differential equations and can be modified for use with other grid based spatial discretization methods.

\subsubsection{A Second-Order Scheme}
A modification of a second-order scheme used by \citet{DonSch11} and implemented in this study is
\begin{align}
&{}\frac{u^{n+1}-2u^n+u^{n-1}}{\delta t^2} -\Delta\frac{u^{n+1}+2u^n+u^{n-1}}{4}  \notag
\\&{} +\frac{u^{n+1}+2u^n+u^{n-1}}{4}=\left\lvert u^n\right\rvert^2u^n, \label{eq:DonSchSemiImplicit}
\end{align}
where $u^n\approx u(n\delta t,x,y,z)$. Time stepping takes place in  Fourier space where the linear elliptic equation from the semi-implicit time discretization is easy to solve, and the three dimensional Fourier transform is used to obtain the nonlinear term in real space. A more detailed explanation of the method can be found in \citet{CloMuiRig12}, \citet{Rig12} and \citet{ParSpec}. Example Matlab and Python implementations of this method, as well as the parallel code can be found in \citet{ParSpec}. This scheme requires two FFTs per time step. The implementation used  in this study allows for the field $u$ to be complex.

\section{Results}
\begin{figure*}[hbt]
\begin{center}
\includegraphics[height=5.5in]{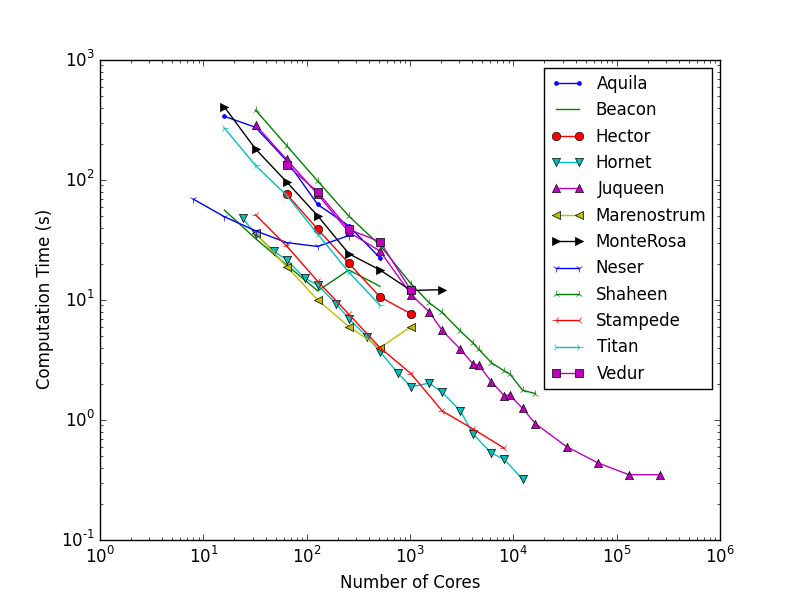}
\caption{Scaling results showing computation time for 30 time steps as a function of the number of processor cores.}
\label{fig:ScalingFigure}
\end{center}
\end{figure*}

We compare the scalability of the numerical scheme, without output to disk\footnote{Except for the VSC2 for which no noticeable difference in runtime was observed when doing output to disk, and when not doing output to disk.}. In all cases the wall clock time for 30 time steps was measured,  taken. Figure \ref{fig:ScalingFigure} shows strong scaling results and Table \ref{table:512ranked} lists the computers according to the shortest run time, as well as documenting the properties of each computer.

\begin{sidewaystable*}[p]
\begin{center}
\caption{A table showing ranking of speed at which different computers solve the Klein-Gordon equation for a grid size of $512^3$. The systems used, their compute chips, interconnect and underlying one dimensional FFT library used by 2DECOMP\&FFT, theoretical chip bandwidth from RAM and theoretical peak floating point performance are also shown. Beacon, Stampede and Titan have accelerators which were not used in the study, hence their properties are not listed nor used in the calculation of theoretical peak double precision floating point performance.}
\label{table:512ranked}

\begin{tabular}{|c|c|c|c|c|c|c|c|c|c|c|}
  \hline
 Rank & Machine & Time & Cores & Manufacturer & Node & Total   & Interconnect & 1D         & Chip           & Theoretical   \\
             & Name     &  (s)    & used   &   and Model    & Type & Cores  &                        &   FFT     & Bandwidth & Peak             \\
             &                 &           &             &                          &           &              &                        & Library & Gb/s             & TFLOP/s       \\
\hline
1 & Hornet & 0.319 & 12,288 & Cray XC40 & 2$\times$12 core Intel Xeon & 94,656 & Cray  & FFTW 3 & 68 & 3,784 \\
 &\citep{Hornet}  &  &  &  &  2.5 GHz E5-2680v3 &  &  Aries & \citep{FFTW3}  &  &  \\
2 & Juqueen&  0.350 & 262,144 & IBM  & 16 core 1.6 GHz   & 458,752 & IBM 5D & ESSL & 42.6 & 5,872  \\
 &\citep{Juqueen}  &   &  &Blue Gene/Q &  Power PC A2  &  &   torus & \citep{ESSL}  &  & \\
3 & Stampede & 0.581 & 8,162 & Dell  & 2$\times$8 core Intel Xeon  & 462,462 & FDR  & Intel MKL & 51.2 & 2,210  \\
 &\citep{Stampede}  &   &  &Power Edge &  2.7 GHz E5-2680  &  & infiniband  & \citep{MKL}  &   & \\
4 & Shaheen & 1.66 & 16,384  & IBM  & 4 core 0.85 GHz  & 65,536 & IBM 3D  & ESSL & 13.6  & 222.8 \\
 &\citep{Shaheen}   &   &  & Blue Gene/P &  PowerPC 450  &  & torus & \citep{ESSL} &     & \\
5 & MareNostrum & 4.00 & 64 & IBM & 2$\times$8 core Intel Xeon  &  48,384 & FDR10  & Intel MKL & 51.2 & 1,017  \\
 & III\citep{Marenostrum}  &   &  & DataPlex  &  2.6 GHz E5-2670  &  &infiniband  & \citep{MKL}  &  & \\
6 & Hector & 7.66 & 1024 & Cray XE6 & 2$\times$16 core AMD Opteron  & 90,112 & Cray & ACML & 85 & 829.0 \\
 & \citep{Hector} &   &  & &  2.3 GHz 6276 16C  &  & Gemini  & \citep{ACML}  &  & \\
7 & VSC2 & 9.03 & 1024 & Megware & 2$\times$8 core AMD Opteron  & 21,024 & QDR & FFTW 3 & 85 & 185.0  \\
 &\citep{VSC2}  &   &  & &  2.2 GHz 6132HE  &  & infiniband  & \citep{FFTW3}  &  & \\
 8 & Beacon & 9.13 & 256 & Appro & 2$\times$8 core Intel Xeon  &768 & FDR  & Intel MKL  & 51.2 & 16.0 \\
 &\citep{Beacon}  &   &  & &  2.6 GHz E5-2670  &  & infiniband &\citep{MKL}  &  & \\
9 & Monte Rosa& 11.9  & 1,024 &Cray XE6 & 2$\times$16 core AMD Opteron  & 47,872 & Cray & ACML & 85 & 402.1\\
 &\citep{MonteRosa}  &   &  & &  2.1 GHz 6272  &  & Gemini  &  \citep{ACML}  &   & \\
10 & Titan & 17.0 & 256 & Cray XK7 & 16 core AMD Opteron & 299,008 & Cray & ACML & 85 & 2,631 \\
 & \citep{Titan} &   &  & &  2.2 GHz 6274   &  & Gemini  & \citep{ACML}  &   & \\
11 & Vedur & 18.6 & 1,024 & HP ProLiant  & 2$\times$16 core AMD Opteron & 2,560 & QDR & FFTW 3 & 85 & 236  \\
 & \citep{Vedur} &   &  & DL165 G7 &  2.3 GHz 6276  &  &  infiniband & \citep{FFTW3}  &   & \\
12 & Aquila & 22.4 & 256 & ClusterVision & 2$\times$4 core Intel Xeon   & 800 & DDR & FFTW 3 & 12.8 & 8.96 \\
 & \citep{Aquila} &   &  & &  2.8 GHz E5462  &  &  infiniband & \citep{FFTW3} &  & \\
13 & Neser & 27.9 & 128 & IBM System  & 2$\times$4 core Intel Xeon & 1,024 & Gigabit &  FFTW 3 & 10.7 & 10.2 \\
 & \citep{Neser} &   &  & X3550 &   2.5 GHz E5420  &  & ethernet  & \citep{FFTW3} &  & \\
  \hline
\end{tabular}
\end{center}

\end{sidewaystable*}

\begin{figure*}[hbt]
\begin{center}
\includegraphics[height=5.5in]{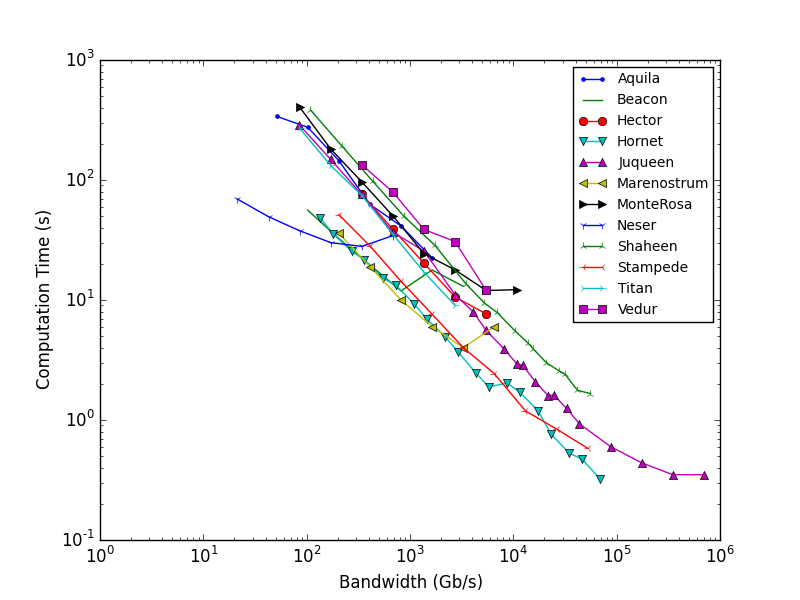}
\caption{Scaling results showing computation time for 30 time steps as a function of total on chip bandwidth defined as the maximum theoretical bandwidth from RAM on a node multiplied by the number of nodes used.}
\label{fig:ScalingFigure2}
\end{center}
\end{figure*}

\subsection{Performance Model}
Before discussing the results, it is helpful to have a model for how fast parallel computation can make this computation. \citet{WilWatPat09} introduced the roofline model which provides a simple upper bound for performance of an algorithm at the node level.  Fast Fourier transforms are typically bandwidth limited, the loop based computations in the code are performed on long vectors with unit stride accesses and little re-use of values loaded from memory, and so are similarly bandwidth limited at the node level. For each Fourier Transform, two MPI\_ALL\_TO\_ALL\_V or MPI\_ALL\_TO\_ALL calls are used. For small message sizes, these are usually latency limited, while for large message sizes, these are usually network bandwidth limited. Since the problem size considered here is not too large, a simple model would use on chip bandwidth and network latency to determine an estimate of performance and scalability. Let a single core have a bandwidth of $B_c$, and the minimum network latency for a single byte be $L_n$. For a discretization of size $N^3$ grid points, each time step requires approximately $d_1\times N^3$ double precision floating point operations and $2d_2\times[N\log(N)]^3$ operations for the FFT where $d_1$ and $d_2$ are constants. Hence, on a single core, the runtime is approximately given by
$\frac{d_1\times N^3+2d_2\times[N\log(N)]^3}{B_c},$
so that on $p$ processes the runtime is
$\frac{d_1\times N^3+2d_2\times[N\log(N)]^3}{B_c\times p},$
assuming no network communication costs. When there is network latency, we get 
$\frac{d_1\times N^3+2d_2\times[N\log(N)]^3}{B_c\times p} + L_n.$
Thus, the network latency gives a lower bound on the possible speedup. It also takes time to send messages across a network. This depends on the topology of the network being used and the algorithm used to perform the MPI\_ALL\_TO\_ALL\_V or MPI\_ALL\_TO\_ALL exchange \citet{BerEtAl91}, \citet[p. 537]{GraGupKarKum03}, \citet{Swa87}, \citet{SwaHam01}.  Since the problem size is not too large, it is reasonable to assume some small network dependent constant $d_3$ multiplied by the natural logarithm of the total number of processes (the lower bound for a hypercube network that is well suited to the FFT) so that we obtain
$\frac{d_1\times N^3+d_2\times[N\log(N)]^3}{B_c\times p} + L_n + d_3\log(p).$
The model is similar though not as detailed as the ones in \citet{Aya13}, \citet{KerBar11} and \citet{KerBarGalCheBruRyuChiHoi13}. Since different runtime FFT algorithms on each machine are used, and since the slab decomposition is used for a small number of process and the pencil decomposition is used for more than 512 processes, and since there are different algorithms for performing an all to all exchange, the best of which will depend on the size of the problem being solved, the computer being used and the number and location of the processors being used on that computer (see \citet{FosWor97}), a more detailed model is not developed. For small $p$ and fixed $N$, the runtime decreases close to linearly, and once $p$ is large enough the runtime starts to increase again due to communication costs. Since most of the computers used in this study are not hypercubes, the model can only provide a lower bound of the time to solution without communication and computation overlap. Given the large number of computers used, and the fact that even on a single computer, process placement, software environment configuration, as well as communications by other compute jobs would affect program performance, it is infeasible to give a more complete model for the communication cost. It should be noted, that such a model may also be applicable to accelerated computers, if there is sufficient network support and a good implementation, see for example \citet{CheCuiMei10}, \citet{CzeEtAl11}, \citet{CzeEtAl12}, \citet{ParEtAl13}, \citet{SonHol14} and \citet{Swa97}.  The model does not account for overlap of computation and communication in computing a distributed Fourier transform, which have been tried by \citet{KanEtAl11} and \citet{HoeZer07} (MPI 3.0 non blocking collectives make this easier to implement), for which one might be able to decrease the lower bound. The work here is similar in spirit to that in \citet{WorFos97}, but the equation used is simpler, and hence easier to use for evaluating computer performance.

\subsection{Result Summary}
Table \ref{table:512ranked} shows that for this problem, a ranking can be obtained that would allow for easy evaluation of computer architectures for solving partial differential equations of a given discretization in the fastest time possible using algorithms that are dominated by Fourier transforms.  Figure \ref{fig:ScalingFigure} shows that the strong scaling limit is not reached on all the platforms primarily due to clusters being too small or due to queue size restrictions. For cases where less cores were used than close to full system size and the strong scaling limit was not reached, further computation time is required, in particular on Titan and Hornet.  It is likely that the results in the table can change with extensive tuning and careful job placement, however they are representative of what occurs in a typical production environment. Figure \ref{fig:ScalingFigure2} shows scaling, but rather than using core count, uses total processor bandwidth which is defined as bandwidth per node multiplied by number of nodes used. \citet{CzeEtAl12} and \citet{MarGarGla14} indicate that node level bandwidth may be more important than system interconnect bandwidth, and indicate that for benchmarks which look at a fixed problem size, a few key performance indicators can replace the benchmark -- for the current code, memory bandwidth and performance of MPI\_ALL\_TO\_ALL are good indicators and for HPCG\citep{HPCG}, \cite{MarGarGla14} indicate that memory bandwidth and MPI\_ALL\_REDUCE time are good performance indicators. Good performance for  MPI\_ALL\_TO\_ALL is much harder to achieve than good performance for  MPI\_ALL\_REDUCE  and will in particular imply good performance for MPI\_ALL\_REDUCE, though likely at a higher hardware cost.

\subsection{Result Discussion}
Table \ref{table:512ranked} shows that Hornet produces the fastest run time due to its high performance communication network and fast processors. Juqueen does however have a much higher theoretical peak floating point performance than Hornet. Juqueen's large number of processor cores are difficult to use efficiently for this problem size, thus despite Hornet's smaller size, it is more effective at solving the Klein Gordon equation using the current algorithm. Similar behavior is observed on Marenostrum III, where network performance makes it difficult to utilize a large portion of the machine despite the high theoretical peak performance. On Aquila, there is a pronounced drop in performance when going from 8 to 16 cores, and after this the scaling is close to ideal again relative to the 16 core run (and the efficiency for 256 cores would be $76.5\%$ if it was measured relative to the 16 core results).  This drop in performance is likely because the 8 core run only requires intra-node communications, whereas all runs with higher core counts have to send messages via the slower infiniband interconnect. A similar drop in performance is observed on Hornet in going from 2048 cores to 3072 cores, since MPI communication requires two communication steps on the dragon fly topology rather than just one communication step. On Beacon, the runtime was quite sensitive to process placement, likely due to network topology. Neser is the oldest machine used in this study. It is a Linux cluster with a 1Gb ethernet network that is still fast at small core counts. Neser's low cost gigabit ethernet network prevents good scaling behavior to larger core counts and explains its low ranking in table \ref{table:512ranked}. Speed up on Shaheen was very close to ideal due to the fast interconnect and balanced design which allows for good throughput to the cores given their maximum floating point performance. Shaheen has 0.85 GHz cores and Juqueen has cores that are clocked at 1.6GHz and can do twice as many floating point operations per cycle,  yet for the same number of cores, Juqueen is on average only 1.4 times faster than Shaheen, this is likely because each core on Juqueen has a lower share of bandwidth (2.66Gb/s) than on Shaheen (3.375 Gb/s). The performance on Vedur is significantly worse than on Hector, despite having the same compute chips. This is likely due to interconnect latency. Finally, fig.\ \ref{fig:ScalingFigure2} also shows that newer Intel processors on Beacon, Hornet, Marenostrum and Stampede give much better performance than the Power PC, older Intel and AMD processors for the same node level bandwidth. The reason for the improved performance is likely to be due to the larger number of floating point operations that can be done per cycle compared to the other architectures, and the more sophisticated memory controllers -- this indicates that simply knowing chip bandwidth and performance of MPI\_ALL\_TO\_ALL allows for a simple but incomplete model. As explained by \citet{LoEtAl14}, these characteristics of a machine are also difficult to measure precisely so their use for predictions should be done with care, though they may give a good but not perfect initial approximation.

\section{Future Work}
The results in this paper give a guide for codes which heavily utilize the Fourier transform on the different combinations of processor and interconnect that will give the best overall computation time. There are a variety of other methods for solving the same equation, and other aspects of using supercomputers that have not been covered in the present study, but would be useful to cover in other studies. Possible future work  includes: finite difference/finite element codes using multigrid, tree code, fast multipole or conjugate gradient linear equation solvers for the implicit linear system solve in the time discretized Klein-Gordon equation, high order timestepping, in-situ visualization, measurement of computer energy consumption, use of accelerators, effectiveness of input and output, and more detailed performance models can be done in further work on a smaller set of computers. Care will be required in choosing initial conditions to allow for a meaningful comparison with other ways of discretizing this equation and solving the linear equations in implicit time stepping schemes.

\subsubsection*{Acknowledgments.} 

The authors thankfully acknowledge the computer resources, technical expertise and assistance provided by:
The Beacon Project at the University of Tennessee supported by the National Science Foundation under Grant Number 1137097;
The UK's national high-performance computing service, which is provided by UoE HPCx Ltd at the University of Edinburgh, Cray Inc and NAG Ltd, and funded by the Office of Science and Technology through EPSRC's High End Computing Programme;
The Barcelona Supercomputing Center - Centro Nacional de Supercomputaci\'{o}n;
The Swiss National Supercomputing Centre (CSCS);
The Texas Advanced Computing Center (TACC) at The University of Texas at Austin;
The KAUST Supercomputer Laboratory (KSL) at King Abdullah University of Science and Technology (KAUST) for providing the resources that have contributed to this study;
The Oak Ridge Leadership Computing Facility at the Oak Ridge National Laboratory, which is supported by the Office of Science of the U.S. Department of Energy under Contract No. DE-AC05-00OR22725;
The Aquila HPC service at the University of Bath;
The Vienna Scientific Cluster (VSC);
The PRACE research infrastructure resources in Germany at HLRS and FZ Juelich;
The High Performance Computing Center of the University of Tartu.
Any opinions, findings, and conclusions or recommendations expressed in this material are those of the authors and do not necessarily reflect the views of the funding bodies or the service providers.

Initial research to start this project used resources of Kraken at the National Institute for Computational Science, Trestles at the San Diego Supercomputing Center both through the Extreme Science and Engineering Discovery Environment (XSEDE), which is supported by National Science Foundation grant number ACI-1053575,  the University of Michigan High Performance Computing Service FLUX and Mira at the Argonne Leadership Computing Facility at Argonne National Laboratory, which is supported by the Office of Science of the U.S. Department of Energy under contract DE-AC02-06CH11357
 
Oleg Batra\u{s}ev and Benson Muite were partially supported  by the Estonian Centre of Excellence in Computer Science (EXCS) under the auspices of the European Regional Development Funds. Mark Van Moer's contribution was made possible through the XSEDE Extended Collaborative Support Service (ECSS) program. We also thank Stefan Andersson, Winfried Auzinger, Andy Bauer, David E DeMarle, David Ketcheson, David Keyes, Robert Krasny, Dmitry Pekurovsky, Michael Pippig, Paul Rigge, Madhusudhanan Srinivasan, Eero Vainikko, Matthias Winkel, Brian Wylie, Hong Yi and Rio Yokota for helpful advice and suggestions.

\bibliography{bibliography.bib}
\bibliographystyle{plainnat} 

\subsection*{Biography}
 
Samar Aseeri (email: \url{samar.asseeri@kaust.edu.sa}) is a computational scientist at the King Abdullah University of Science and Technology.

Oleg Batra\u{s}ev (email: \url{olegus@ut.ee}) is a researcher in computer science at Tartu \"{U}likool.

Matteo Icardi (email: \url{matteo.icardi@kaust.edu.sa}) is a postdoctoral research fellow at the King Abdullah University of Science and Technology.

Brian Leu (email: \url{brianleu@umich.edu}) is an undergraduate student in electrical engineering at the University of Michigan. 

Albert Liu (email: \url{alberliu@umich.edu}) is an undergraduate student in electrical engineering and physics at the University of Michigan.

Ning Li (email: \url{ning.li@nag.co.uk}) is a high performance computing consultant at the Numerical Algorithms Group. Ning specializes in numerical software development, in particular in the area of high performance computing and wrote the parallel software framework 2DECOMP\&FFT.

Benson Muite (email: \url{benson.muite@ut.ee}) is a postdoctoral research fellow in computer science at Tartu \"{U}likool.

Eike M\"{u}ller (email: \url{e.mueller@bath.ac.uk}) is a postdoctoral research associate in mathematics at the University of Bath.

Brock Palen (email: \url{brockp@umich.edu}) is a high performance computing system administrator at the University of Michigan.

Michael Quell (email: \url{michael.quell@yahoo.de}) is an undergraduate student in mathematics at the Technische Universit\"{a}t Wien.

Harald Servat (email: \url{harald.servat@bsc.es}) is a computer science researcher at the Barcelona Supercomputing Center.

Parth Sheth (email: \url{pssheth@umich.edu}) is a research assistant in biomedical engineering at the University of Michigan.

Robert Speck (email: \url{r.speck@fz-juelich.de}) is a mathematician at Forschungszentrum Juelich, Juelich Supercomputing
Centre.

Mark Van Moer (email: \url{mvanmoer@illinois.edu})  is a visualization scientist at the National Center for Supercomputing Applications.

Jerome Vienne (email: \url{viennej@tacc.utexas.edu}) is a computational scientist at the Texas Advanced Computational Center.

\end{document}